\documentclass[manuscript,sigconf]{acmart}

\usepackage{balance}
\usepackage{tikz}

\usepackage{svg}
\AtBeginDocument{%
  \providecommand\BibTeX{{%
    \normalfont B\kern-0.5em{\scshape i\kern-0.25em b}\kern-0.8em\TeX}}}

\setcopyright{none}

\acmConference[ARES 2024]{The 19th International Conference on Availability, Reliability and Security}{July 30-August 2, 2024}{Vienna, Austria}
\acmBooktitle{The 19th International Conference on Availability, Reliability and Security (ARES 2024), July 30-August 2, 2024, Vienna, Austria}
\acmISBN{979-8-4007-1718-5/24/07}

\graphicspath{{./images/}}
\begin{document}

\title{RMF: A Risk Measurement Framework for Machine Learning Models}


\author{Jan Schröder}
\email{schroeder.jan@protonmail.com}
\orcid{0009-0007-1222-0747}
\affiliation{
  \institution{Fraunhofer FOKUS, HTW Berlin}
  \city{Berlin}
  \country{Germany}
}

\author{Jakub Breier}
\email{jbreier@jbreier.com}
\orcid{0000-0002-7844-5267}
\affiliation{
  \institution{TTControl GmbH}
  \city{Vienna}
  \country{Austria}
}


\begin{abstract}
Machine learning (ML) models are used in many safety- and security-critical applications nowadays. It is therefore important to measure the security of a system that uses ML as a component. This paper focuses on the field of ML, particularly the security of autonomous vehicles. For this purpose, a technical framework will be described, implemented, and evaluated in a case study. Based on ISO/IEC 27004:2016, risk indicators are utilized to measure and evaluate the extent of damage and the effort required by an attacker. It is not possible, however, to determine a single risk value that represents the attacker's effort. Therefore, four different values must be interpreted individually.
\end{abstract}




\maketitle

\section{Introduction}
Machine learning (ML), especially deep learning, is currently one of the most popular subjects in computer science. As the popularity of ML increases, so does the risk of \textit{adversarial machine learning}~\cite{lowd2005adversarial} — the risk of neural networks (NN) being influenced maliciously. The security of ML is especially relevant in safety-critical areas such as medical imaging diagnosis, automated driving, and connected and cooperative mobile machinery.

In this paper, we focus on measuring poisoning attacks that misclassify the labels of images in NN and the effort required by the attacker to execute the attack. We use the term "measurements" to refer to the recording of the process, detailing what an attacker needs to do and what resources they need to perform their specific attack. For a better understanding of the measurement process, a case study demonstrates a concrete example involving the classification of road signs for autonomous vehicles. This measurement computes the risk involved with a poisoning attack and evaluates the effort required by an attacker.

The measurement of poisoning attacks leads to the following questions: \textit{How much damage is possible with an attack against a NN? What is the minimum effort an attacker needs to invest to achieve predefined damage?} The objective of this paper is to devise a framework for quantifying the extent of damage and the attacker's effort in order to derive a risk value that describes the level of risk associated with adversarial attacks against NN. This risk level is intended to describe how significant the risk is when an attack is executed, depending on whether the attacker has a low or high effort. The proposed framework should take into account the requirements specified by the ISO/IEC 27004:2016 standard as accurately as possible for measuring information security risk.

\noindent\textbf{Our Contribution.} This paper follows the requirements of ISO/IEC 27004:2016 \cite{ISO_27004_2016} and provides the following contributions:
    \begin{enumerate}
        \item We propose quantitative and qualitative attributes that should be instantiated with data during measurement methods.
        \item We develop measurement functions to compute values representing the extent of damage and the attacker's effort.
        \item We evaluate the computed values and interpret the final risk of the attack.
    \end{enumerate}
This paper demonstrates how to measure risks to improve security performance. It does not propose countermeasures. The focus of this paper is on NN and poisoning attacks, especially backdoor attacks that misclassify labels of images during inference time.

\noindent\textbf{Paper Organization.} This paper is organized as follows: Section \ref{2} describes the state of the art related to this paper. Section \ref{3} declares the proposed attributes and explains the Threat Model utilized in this paper. Section \ref{sec:rmf} outlines the framework based on ISO/IEC 27004:2016 and discusses the problems encountered while evaluating the measurement process. Section \ref{5} evaluates the framework and the collected data, highlighting the issues that arise after the measurement. Section \ref{6} summarizes the paper and suggests possible future work.

\section{Background and Related Work}
\label{2}

\subsection{Adversarial Attacks on Neural Networks}

Adversarial attacks on ML models pose a significant threat to the security and privacy of entities relying on them. Attackers could exploit the vulnerabilities of ML models to access sensitive data or manipulate the model's output, leading to potential harm to individuals or organizations. In the context of ML security, attacks to consider include data poisoning~\cite{mei2015using}, which we utilize in this paper, evasion~\cite{DBLP:journals/comsur/WangKBDCZ24}, model stealing~\cite{papernot2017practical}, and membership inference attacks~\cite{shokri2017membership}.

Backdoor Attacks can be divided into two broad categories: poisoning-based and non-poisoning-based attacks~\cite{li2022backdoor}.

Poisoning attacks are used to manipulate training data to take control of the ML model \cite{DBLP:conf/iclr/GeipingFHCT0G21}. This can occur, for example, if datasets made available to the public (e.g., on the Internet) already contain poisoned data. If these datasets are not thoroughly checked for vulnerabilities, backdoors can be implemented in images without any problems, potentially allowing label flipping once triggered. To show how the metrics change, we used the Adversarial Robustness Toolbox (ART) \cite{DBLP:journals/corr/abs-1807-01069}, which is described later in this section.

Papers such as \cite{YERLIKAYA2022118101,DBLP:conf/sp/JagielskiOBLNL18} show examples of poisoning attacks against ML models with the goal of misclassifying images by imprinting specific target or random labels during training. To make these attacks successful, clean-label \cite{turner2018clean} and hidden-trigger \cite{DBLP:conf/aaai/SahaSP20} backdoor attacks are used to implement backdoors with optimization techniques. The work of Gu et al. \cite{DBLP:journals/access/GuLDG19} demonstrates that simple changes to an image can completely alter its classification. For example, the authors showed that a stop sign could be predicted as a speed limit sign using a simple pixel pattern in poisoned training data, reducing accuracy by 25\%. This example highlights the need to improve security performance for ML models. The consequences for public safety, such as in traffic due to autonomous vehicles, could otherwise be tragic. To identify these vulnerabilities during the training of an ML model, we show how the risk of an attack can be measured during training and inference.

\noindent\textbf{Non-Poisoning-Based Attacks.} This type of backdoor attack does not tamper with the training data to embed the backdoor into the model but rather changes the model itself~\cite{dumford2020backdooring}.
One direction is to utilize physical attack vectors~\cite{breier2022practical}, such as Rowhammer~\cite{mutlu2019rowhammer}, to flip the bits in the memory where the model parameters are stored~\cite{rakin2020tbt}.
Another direction is to tamper with the training process so that some of the executed processor instructions are skipped, resulting in incorrect evaluation of activation functions~\cite{breier2022foobar}.
Without a backdoor, this technique can also be used during the inference phaset to incur non-targeted evasion~\cite{breier2018practical}.

Another branch of non-poisoning attacks are structure-modified backdoor attacks that change the model structure of benign models.
This can be done by inserting a small number of malicious neurons into the already trained network~\cite{tang2020embarrassingly}.
Another approach simply replaces the original model by the backdoored model, given that the attacker has access rights~\cite{li2021deeppayload}.

\subsection{Adversarial Robustness Toolbox} 
The adversarial attacks and countermeasures can be implemented with the help of the Adversarial Robustness Toolbox (ART).
Various backdoor attacks are available within ART, based on studies such as \cite{DBLP:journals/access/GuLDG19,turner2018clean}. 
These are implemented by specifying parameters such as the number of training and test data to be poisoned, labels to be poisoned, and which label should be used as a target for misclassification. 
Once the attack has been successfully executed using ART, the framework can return the effect of the attack depicted as ML metrics. It should be noted that the metrics are calculated according to the robustness measures described in \cite{DBLP:conf/iclr/WengZCYSGHD18,DBLP:conf/icml/ArpitJBKBKMFCBL17,DBLP:conf/cvpr/Moosavi-Dezfooli16}. This means that the metrics can then be output to show the differences between benign and poisoned data as Table \ref{tab1} shows for the case study. Afterward, ISO/IEC 27004:2016 \cite{ISO_27004_2016} can be consulted to ensure that these results are measured and classified correctly.

In order to measure the risk of backdoor attacks, we proceed according to the requirements from ISO/IEC 27004:2016 which is described in the following section.

\subsection{Risk Management for Machine Learning}

With ISO/IEC 27004:2016, the goal is to evaluate and improve the performance of information security through risk measurement \cite{ISO_27004_2016}. Its focus lies on the measurement, analysis, and evaluation of an organization's system and its information security management system (ISMS). The standard specifies a measurement process designed to fulfill these requirements.

The first step in the measurement process is to specify the attributes, which are determined by the information needs of stakeholders. Measurement methods are then used to measure each attribute. The output of these measurement methods is the base measures, with each base measure corresponding to an attribute. The next step involves computing derived measures using measurement functions to analyze the data. Measurement functions combine two or more base measures (e.g., by computing the average value or adding them together). The analytical model combines the derived measures and decision criteria to compute a value called the indicator. This indicator is used to interpret the measurement results and suggest possible changes to improve security performance in relation to the information needs.

\section{Definition of Attributes for the Risk Measurement}
\label{3}

Breier et al. \cite{DBLP:journals/corr/abs-2012-04884} proposed a risk management framework tailored for organizations that utilize NN for their business use cases. They argued that with the introduction of NN, specific risks need to be taken into account. They suggested using the attacker's knowledge and goal, attack time, and specificity to classify attacks. In addition to these attributes, accuracy is repeatedly used as a metric to represent the amount of damage from backdoor attacks \cite{DBLP:journals/access/GuLDG19}. \\

In this paper, we propose the following additional attributes to measure risk:

\begin{itemize}
    \item \textbf{Computational Resources:} The minimum hardware resources an attacker needs to develop and execute their attack.
    \item \textbf{Attacker's Knowledge, Attacker's Goal, Attack Time, and Attack Specificity:} Attributes based on the work of Breier et al., which are further explained in Table \ref{tab:attributes}.
\end{itemize}

\subsection{Threat Model}

Doynikova et al. \cite{DBLP:conf/crisis/DoynikovaNGK20} proposed a mapping process to differentiate collected attack data and map it to measure the attacker's effort. The mapping process is structured as follows: first, the attributes are derived at high-level and low-level. High-level data cannot be derived directly from the monitored raw data. All data that can be derived directly from the raw data are assigned to the low-level attributes. The core process in this threat model is the mapping between the low- and high-level attributes. Doynikova et al. describe that all high-level attributes represent the attacker, while low-level attributes are monitored data during the attack. Finally, this mapping should provide the ability to objectively measure an attacker's effort. For example, we represent the attacker's knowledge by detailing the steps required, such as whether it is a blackbox or whitebox attack, which indicates what the attacker needs to know to execute the attack on a NN.

\section{Risk Measurement Framework based on ISO/IEC 27004:2016}
\label{sec:rmf}

This section elaborates on the concept of the Risk Measurement Framework (RMF), developed to measure the risk of adversarial attacks against NN by quantifying the impact of the attacks on the performance metrics of the models. The architecture of the RMF is designed based on the requirements of the ISO/IEC 27004:2016 standard, which aims to establish a systematic approach to the measurement of information security risks. The framework comprises several components, including a data collector and a risk evaluator, whose goal is to calculate the risk value.

\subsection{Data}
\label{sec:data}

The implemented framework acquires data from a tested NN. The effort of an attacker is measured based on their knowledge, goal, attack specificity (all the three parameters are monitored in terms of steps \cite{bsi_2013}), the computational resources at their disposal (CPU, GPU, memory), and the attack time. A step is a non-negative integer that represents a sub-goal the attacker must achieve to execute the attack successfully. We developed the steps based on the implemented attacks of ART which are which in turn are implemented with the works of \cite{DBLP:journals/access/GuLDG19,turner2018clean} among others. For example, the steps include tasks such as choosing a backdoor, implementing an optimization function, train the original dataset and a proxy classifier. The extent of damage is evaluated through metrics such as F1-Score, average precision, average recall, and accuracy, which are the resulting output of the ART framework attacks.

\subsection{Architecture}

The framework is designed based on the requirements of ISO/IEC 27004:2016 \cite{ISO_27004_2016}, which means the declaration of objects and attributes is the first part. As stated in Section \ref{sec:data}, the effort of the attacker and the extent of damage are measured separately.

\noindent\textbf{Attributes.} The following attributes used in this framework were proposed and adopted from the work of Breier et al. \cite{DBLP:journals/corr/abs-2012-04884} — the knowledge and goal of an attacker, specificity, and time of an attack. Accuracy (after the attack) is a common adversarial ML metric used to depict the level of successful misclassification by a backdoor attack \cite{DBLP:journals/access/GuLDG19}. Computational resources is an attribute newly included in this work. Every attribute is mapped to a measurement method that measures the training and inference time of the original and manipulated NN. These attributes are listed in Table~\ref{tab:attributes}.

The attacker's knowledge is an attribute that represents everything the attacker needs to know about the NN. The attacker's goal depicts what the attacker needs to do to achieve their goal based on the information of the attacker's knowledge. Attack specificity consists of the steps required to implement an attack that changes a label to a specific or random label. This specificity is particularly relevant for backdoor attacks because a backdoor attack can be targeted or untargeted. For attack time, the framework measures the interval of how long the attack takes, including the training and inference time. Due to the indistinguishability of the original and manipulated NN with respect to the three identified attributes, a comparative analysis with other attacks and their objectives is required for differentiation. This analysis was performed by executing two different attacks on one NN. In this work, we were able to distinguish between the original and manipulated NN by measuring differences in computational resources, accuracy, F1-Score, average precision, and average recall.

\begin{table}
    \centering
    \caption{Attributes used in the proposed RMF and their description.}
    \label{tab:attributes}
    \begin{tabular}{p{3.5cm} p{4cm}}
        \toprule
       Attribute  & Description \\ \midrule
       Attacker's knowledge & White-box or black-box attack described in steps. \\
       Attacker's goal & Steps to achieve the goal with the executed attack. \\
       Attack time & Time in seconds to measure the training with the executed attack. \\
       Attack specificity & Steps to execute an attack on a specific or random image label. \\
       Computational resources & Minimum hardware resources to execute the attack. \\
       Accuracy & ML metric to show the difference between the original and poisoned dataset. \\
       F1-Score, Average Precision, and Average Recall & Attribute to show the difference between the original and poisoned dataset for each image. \\ 
       \bottomrule
    \end{tabular}
\end{table}

\noindent\textbf{Measurement.} The next step in the architecture is the summarization of the values to determine the extent of damage and the attacker's effort. ISO/IEC 27004:2016 specifies this summarization with measurement functions (e.g., base measures calculated into derived measures). This process of calculating relevant attributes for the extent of damage and the attacker's effort was conducted through a threat modeling approach, where the measured values monitored during the attack were mapped to the attacker's effort \cite{DBLP:conf/crisis/DoynikovaNGK20}. This mapping process was further translated into measurement functions to quantify the attacker's effort in terms of computational resources, attack time, and the number of steps taken by the attacker.

In the framework, average precision, average recall, and F1-Score were calculated based on the results of the F1-Score, average precision, and average recall. The composite extent of damage is computed by integrating the values of average precision, average recall, F1-Score, and accuracy. The composite attacker's effort is computed by integrating the values of attack time, computational resources, and the number of steps taken by the attacker.

To compute these values, it is necessary to unify them into non-negative numbers. This unification for the extent of damage starts with reversing the poisoned value, as the goal of an attack is to decrease the metric values as much as possible. Since accuracy and other ML metrics are represented as percentages and thus lie between 0 and 100\% (or 0 and 1), the RMF reverses these values. For example, if the accuracy is 0.06 (or 6\%), the RMF converts this value to 0.94 out of 1. This allows the value to appropriately reflect how much the accuracy decreases the overall risk.

By applying this calculation, it can be verified that the damage severity decreases as the poisoned value decreases. This relationship can be observed through the composite metric, which captures the impact of the poisoned value on the system's overall performance. As the poisoned value decreases, the system's ability to accurately predict outcomes improves, leading to a reduction in the overall damage severity. This relationship can be quantitatively assessed through the composite metric, providing a robust means of evaluating the impact of the poisoned value on the system.

As of now, it is not possible to standardize the attacker's effort values because there is no consistent metric that accounts for the variability in attack time (in seconds), computational resources (hardware usage in percent and MB), and attack steps (non-negative natural number). Unlike the poisoned value, which has a direct and measurable impact on the system's overall performance, the attacker's effort values are influenced by a range of complex and interrelated factors that cannot be easily quantified or normalized. Therefore, the extent of damage and the values for the attacker's effort must be considered separately in the results. The metrics are listed in Table~\ref{tab:comp_vals}.

\begin{table}
    \centering
    \caption{Proposed metrics to show the risk of an attack in the RMF.}
    \label{tab:comp_vals}
    \begin{tabular}{lp{4cm}} \toprule
       Metric  & Description \\ 
       \midrule
       Attack steps & All steps added together to show the effort in steps. \\
       Attack time & Time an attacker needs to execute their attack together with the training and inference time. \\
       Computational resources & Minimum hardware resources an attacker needs to execute their attack. \\
       Extent of damage & Computed damage after executing the attack. \\
        \bottomrule
    \end{tabular}
\end{table}

\begin{figure}[!h]
    \begin{center}
        \includegraphics[width=0.47\textwidth]{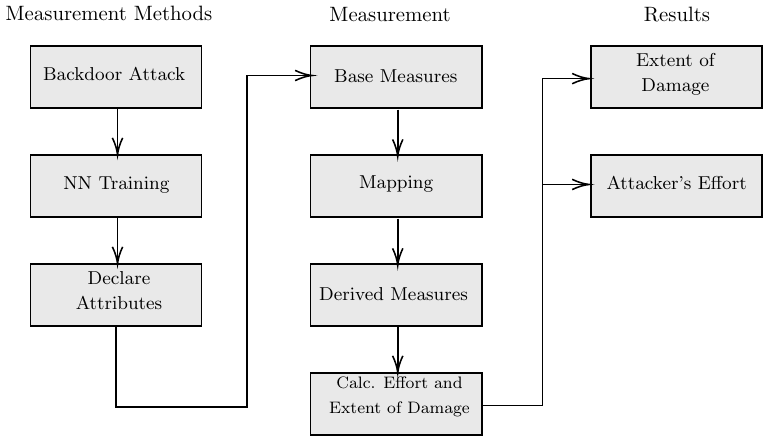}
        \caption{Complete architecture of the RMF with a correctly placed NN during the measurement process.} \label{fig2}
    \end{center}
\end{figure}

\subsection{Attack Implementation}

\begin{figure}[!h]
    \begin{center}
        \includegraphics[width=0.3\textwidth]{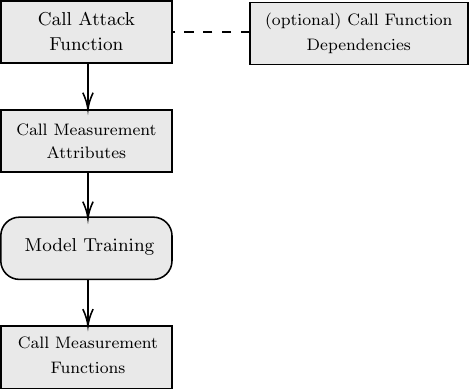}
        \caption{Flowchart on how to utilize the framework in a hierarchical order.} \label{fig10}
    \end{center}
\end{figure}

Figure \ref{fig10} shows how to utilize the framework, which is implemented in Python using the ART Python package. First, the user selects an attack for the measurement by calling the appropriate function and its necessary dependencies. Then, the user chooses the attributes to be measured and calls the measurement methods after the training part in the NN. In the final step, the functions to compute and evaluate the monitored data are utilized. The implemented backdoor attacks are based on the work of Gu et al. \cite{DBLP:journals/access/GuLDG19} and Turner et al. \cite{turner2018clean}.

\subsection{Case Study: Street Sign Detection}

\begin{figure}[!h]
    \begin{center}
        \includegraphics[width=6.5cm]{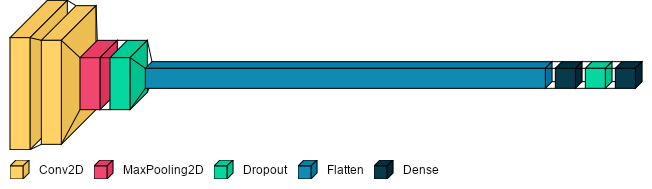}
        \caption{Layered view of the described NN from Table \ref{tab3} without the input layer.} 
        \label{fig9}
    \end{center}
\end{figure}

A case study was conducted where the RMF was used to measure the extent of an attack on an example NN shown in Figure \ref{fig9}. The NN was trained using a dataset of German street signs, as described in \cite{DBLP:conf/ijcnn/StallkampSSI11}, which contains 133,000 images with 70 labels. These labels are clustered into categories such as Speed limit, Prohibitory, Derestriction, Mandatory, Danger, and Unique signs. Furthermore, the NN lacks countermeasures to assess the maximum possible extent of damage and the minimal effort required by an attacker. The goal is also to compare the training and inference performance before and after a successful attack. Due to the lack of data, it is not possible to compare the attack with NNs that have implemented countermeasures or different types of attacks.

\begin{table}
    \begin{center}   
    \caption{NN architecture generated with the Keras Python package.}\label{tab3}
        \begin{tabular}{llr}\toprule
            Activation function &  Keras layer & Shape \\
            \midrule
            None & Input Layer & 30, 30, 3 \\
            ReLU & Convolutional 2D & 28, 28, 32\\
            ReLU & Convolutional 2D & 26, 26, 64\\
            None &  Max Pooling 2D & 13, 13, 64\\
            None & Dropout & 13, 13, 64 \\
            None & Flatten & 10816 \\
            ReLU & Dense & 128\\
            None & Dropout & 128\\
            Softmax & Dense (Output Layer) & 43 \\
            \bottomrule
        \end{tabular}
    \end{center}
\end{table}

The computer used to train the model and execute the attack is equipped with an AMD Ryzen 7 5800X 8-Core Processor running at 3.80 GHz, 16 GB of RAM, and an NVIDIA GeForce RTX 3060 Ti graphics card. Table \ref{tab1} shows the data measured with the original dataset and the poisoned dataset. We execute the clean-label backdoor attack \cite{turner2018clean} by poisoning 50\% of the training dataset on random images and classify all of them to label 10. As already mentioned in Section \ref{2}, it is possible to output the ML metrics for successful attacks with the ART. This makes it possible to show the risk posed by an attack and to compare how much the results deviate from the original metrics. Table \ref{tab2} shows the data measured during the attack without references to the original values.

\begin{table}
    \begin{center}   
    \caption{Measured data once with the original dataset and once with the poisoned dataset.}\label{tab1}
        \begin{tabular}{lcc}
            \toprule
            Attribute &  Original value & Poisoned value\\
            \midrule
            Accuracy & 0.94 & 0.06 \\
            Average Precision & 0.96 & 0.02 \\
            Average Recall & 0.94 &  0.02 \\
            F1-Score & 0.94 & 0.01 \\
            \bottomrule
        \end{tabular}
    \end{center}
\end{table}

\begin{table}
    \begin{center}
    \caption{Measured data during the attack without references to original values.}\label{tab2}
        \begin{tabular}{lp{4cm}}
            \toprule
            Attribute &  Clean label backdoor \\
            \midrule
            Computational resources & 9\% CPU, 8137.81MB GPU \\
            Attacker's knowledge & 10 steps \\
            Attacker's goal & 6 steps \\
            Attack time & 1037.23 seconds \\
            Attack specificity & 5 steps \\ 
            \bottomrule
        \end{tabular}
    \end{center}
\end{table}

In summary, the following four values were the result of the measurement in the case study: 4.62 is the extent of damage. This value is calculated by inverting and adding the poisoned ML metrics. 21 attack steps originate by adding the knowledge, goal, and specificity together to get the overall steps that are needed to implement and execute the attack. 1037.23 seconds required for the NN to train along with the attack execution, and the computational resources used were 9\% CPU usage and 8137.81 MB of GPU RAM usage. In the next section, these results and the measurement process will be evaluated and interpreted. Based on the risk analysis standard from the European Telecommunication Standards Institute (ETSI) \cite{applications_2022}, this would be considered a critical risk as shown in Table \ref{tab:risk}. For road sign recognition in an autonomous vehicle, these results mean that each image with a backdoor gets a different label predicted, and thus, for example, a stop sign could be misclassified as a speed limit sign.

\begin{table}
    \begin{center}
    \caption{Classification of risks that is standardized by ETSI \cite{applications_2022}.}\label{tab:risk}
        \begin{tabular}{lp{5cm}}
            \toprule
            Risk &  Description \\
            \midrule
            Minor & The probability of an attack is low and there is no need for action. \\
            Major & These risks should be handled seriously and be minimized with countermeasures. \\
            Critical & Critical risks should be minimized as fast as possible with appropriate countermeasures. \\
            \bottomrule
        \end{tabular}
    \end{center}
\end{table}

\section{Evaluation}
\label{5}

The case study demonstrates the ability to measure the extent of damage based on the used attributes and the effort an attacker must expend, as per the requirements of ISO/IEC 27004:2016. Based on the collected data, the average precision, average recall, accuracy, and F1-Score represent the extent of a backdoor attack on a NN because the ART returns the differences between original and attacked NN.

We decided to invert the extent of damage result to demonstrate that the closer the results from the attack are to the lowest risk value, the better the security performance of the NN is. At the same time, higher effort values should represent a lower risk for the NN, indicating that the attack is becoming increasingly difficult.

Due to the lack of comparable data, it is not possible to present the risk in a plausible way. One way to compare the extent of damage is to utilize the individual data from the NN training with the original dataset. This makes it possible to compare how far apart the results are. If one uses the same calculation as for the calculation of the damage extent, then it comes out that with the original training dataset, 0.1 is calculated. This difference shows a possible goal to reduce the extent of the damage. If there is enough comparable data, a risk matrix could be created \cite{applications_2022}. For the attacker's effort, this is not possible because there is no comparable data that represents the effort for other attacks.

\section{Conclusion}
\label{6}

This paper shows that a risk measurement can show how much damage an attack can do on a NN and what an attacker has to do to perform the attack successfully. Performing this measurement process based on ISO/IEC 27004:2016 \cite{ISO_27004_2016} during the development of a NN helps to improve the security performance to prevent possible attacks before deployment. \\ The results of the case study have shown that the misclassification represents a critical risk. This can be determined by the fact that no countermeasures were taken, which allows for the highest possible damage with a low effort for the attacker. \\ In the future if more attacks are measured in this way, then it will be possible to compare how easily an attacker can carry out an attack and what damage he can achieve with the corresponding effort. Adding the required effort for an attacker to the risk measurement makes it possible to increase the required resources an attacker has to utilize. It is therefore important that more attacks are measured in the future to implement more specific countermeasures in the development of a NN. 

\vspace{0.5cm}
\noindent
\textbf{Acknowledgment.}
This research was funded by the European Commission, under the Horizon Europe project aerOS, grant number 101069732.

\balance
\bibliographystyle{ACM-Reference-Format}
\bibliography{sample-base}

\end{document}